\documentclass[letterpaper]{jpconf}
\usepackage{graphicx}
\begin{document}
\title{Upsilon Productions at STAR}

\author{A. M. Hamed for the STAR Collaboration}

\address{Texas A\&M University}

\begin{abstract}
The $\Upsilon(1S+2S+3S)\rightarrow e^{+}e^{-}$ cross section is measured at mid-rapidity ($y$) in $p+p$ 
collisions and in d$+Au$ collisions at center-of-mass energy $\sqrt{s}$ = 200 GeV with the STAR detector at RHIC. In $p+p$, 
the measured cross section is found to 
be consistent with the world data trend as a function of $\sqrt{s}$,
in agreement with the Color Evaportaion Model (CEM), and underestimated by the Color Singlet Model (CSM) up to the
Next-to-Leading-Order Quantum Chromodynamics (NLO QCD) calculations. In d+$Au$, the measured
cross section is in agreement with the CEM prediction with anti-shadowing effects, and 
the nuclear modification factor indicates that $\Upsilon(1S+2S+3S)$ production follows binary scaling within the current uncertainties.
These measurements provide a benchmark for the future measurements of $\Upsilon$ production in $Au+Au$ collisions.
\end{abstract}

\section{Introduction \label{sec:intro}}
The measurement of heavy quarkonia ($J/\Psi$, $\chi_{c}$, $\Upsilon$, etc) 
is considered as one of the most promising probes of 
the formed medium at the Relativistic Heavy Ion Collider (RHIC). The force between the constituents of a quarkonium state, a heavy quark and its
antiquark, is weakened by the color screening produced by the light quarks and gluons~\cite{Matsui}. Accordingly the quarkonium
yields are expected to suppress in $Au+Au$ collisions compared to those in $p+p$ collisions.
Recent calculations based on Lattice QCD predict a sequential disappearance of the
heavy quarkonia and its excited states 
with the temperature of the strongly interacting systems.
Therefore, the heavy quarkonia and its excited states are considered as a thermometer for
such system~\cite{thermo}. In particular, the measurement of $\Upsilon(1S+2S+3S)$ is recognized as a cleaner probe 
due to its low cross section where the roles of the competing effects that either
reduce the yield~\cite{co-mover}, or enhance it~\cite{recomb}, are negligible in contrast to the $J/\psi$ and its excited states. Furthermore, the contributions
from the higher states ``feed-down" of $b\bar{b}$ is small compared to the $c\bar{c}$ states, where the B meson decays have significant
contributions. Since the $p+p$ collisions serve as a baseline for the $Au+Au$ collisions, it is important to understand
the quarkonia production mechanisms in $p+p$, where almost all of the models have difficulties to reproduce the cross sections,
transverse momentum p$_{\perp}$ spectra, and polarization simultaneously~\cite{pp-models}.

In this article, we report the STAR measurement of the $\Upsilon$ cross section at midrapidity in $p+p$ collisions at 
$\sqrt{s}$ = 200 GeV. In addition, we address the cold nuclear matter effect by performing similar measurements in d$+Au$ collisions at the
same $\sqrt{s}$.  

\section{Data and Analysis}
The STAR experiment is adapted for the quarkonium measurements, through the dielectron channels, 
due to the capabilities of the Time Projection Chamber (TPC)~\cite{TPC} 
and the Barrel Electromagnetic Calorimeter (BEMC)~\cite{BEMC} for electron identification and triggering. 
The TPC and BEMC have a
large acceptance in pseudorapidity $|\eta| < $ 1 and full coverage in azimuth $\phi$. 
The TPC is a tracker detector that measures the particles' momenta and energy loss per unit length 
(dE/dx) and therefore provides particle identification.
The BEMC is divided in $\eta \times \phi$ into
4800 projective towers. 
The geometrical acceptance of the STAR BEMC for detecting
both electrons from an $\Upsilon$ decay is maximum
in the rapidity range $|y| < $ 0.5.

The STAR $\Upsilon$ trigger
is composed of two components, level-0 (L0) hardware component and level-2 (L2) software component~\cite{STAR-ppupsilon}.
Using a dedicated trigger algorithm  
exploiting the capabilities of the STAR BEMC, the STAR
experiment sampled an integrated luminosity of 7.9 pb$^{-1}$ of $p+p$ collisions and 32 nb$^{-1}$ of 
d$+Au$ collisions at $\sqrt{s}$ = 200 GeV.

The electrons identified by the TPC are further required to match the trigger conditions at the BEMC, and to have 
the ratio of E/p $\sim$ 1, where
E and p are the particle's energy and momentum measured by BEMC and TPC respectively. 
The $e^{+}e^{-}$ pairs are then combined to produce the invariant mass ($M_{ee}=\sqrt{2p_{1}p_{2}(1-\cos\theta)}$ where $p_{1}$ 
and $p_{2}$ are
the pair's momenta, and $\theta$ is the opening angle) spectrum
($N_{+-})$, and 
the like-sign combinations of $e^{+}e^{+}$ and $e^{-}e^{-}$ are used to
calculate the geometric mean (2 $\sqrt{N_{++}N_{--}}$) for the combinatorial background. The total yield ($N$ =
$N_{+-}$ - 2 $\sqrt{N_{++}N_{--}}$) is extracted
by integrating the invariant mass spectrum in the $\Upsilon(1S+2S+3S)$ region after the background
subtraction. A detailed study of the line shape is performed in order to separate the $\Upsilon(1S+2S+3S)$ from the
continuum contributions (Drell-Yan and $b\bar{b}\rightarrow e^{+}e^{-}$) in the integrated yield region~\cite{STAR-ppupsilon}.
In these analyses the separation between the individual states of $\Upsilon$ is not possible due the limited statistics 
and Bremsstrahlung tails. 
The cross section is obtained from the total yield $N$ according to equation 1.
\begin{equation}\label{xsection}
\sum_{n=1}^3{\cal{B}}(nS)\times \sigma(nS)={{N}\over{\Delta y\times\epsilon\times{\cal{L}}}} \hspace{1mm},
\end{equation} 
where ${\cal{B}} (nS)$ is the branching fraction for $\Upsilon(nS)\rightarrow e^{+}e^{-}$, $\sigma (nS)$ is the cross section 
$d\sigma/dy$ for the $nS$ state in the region $|y_{\Upsilon}| \le$ 0.5, 
$\Delta y$ is the rapidity interval for our our kinematic region $|y_{\Upsilon}| \le$ 0.5, $\epsilon$ is the total efficiency
corrections, and ${\cal{L}}$ is the integrated luminosity. The total efficiency corrections is defined as 
$\epsilon$= $\epsilon_{geo}\times\epsilon_{vertex}\times\epsilon_{L0}\times\epsilon_{L2}\times\epsilon_{TPC}\times\epsilon_{R}\times\epsilon_{dE/dx}
\times\epsilon_{E/p}$, where $\epsilon_{geo}$ is the BEMC geometrical acceptance, $\epsilon_{vertex}$ is the vertex-finding efficiency,
$\epsilon_{L0}$ is the trigger efficiency for L0, $\epsilon_{L2}$ is the trigger efficiency for L2, $\epsilon_{TPC}$ is the TPC geometrical acceptance times tracking
efficiency, $\epsilon_{R}$ is the TPC-BEMC $\eta-\phi$ matching efficiency, $\epsilon_{dE/dx}$ and $\epsilon_{E/p}$ are the electron
identification efficiency. Each term and its uncertainty is estimated through simulations and the total efficiency is included in the reported values
of the cross section.
\begin{figure}
\begin{center}
   \resizebox{60mm}{!}{\includegraphics{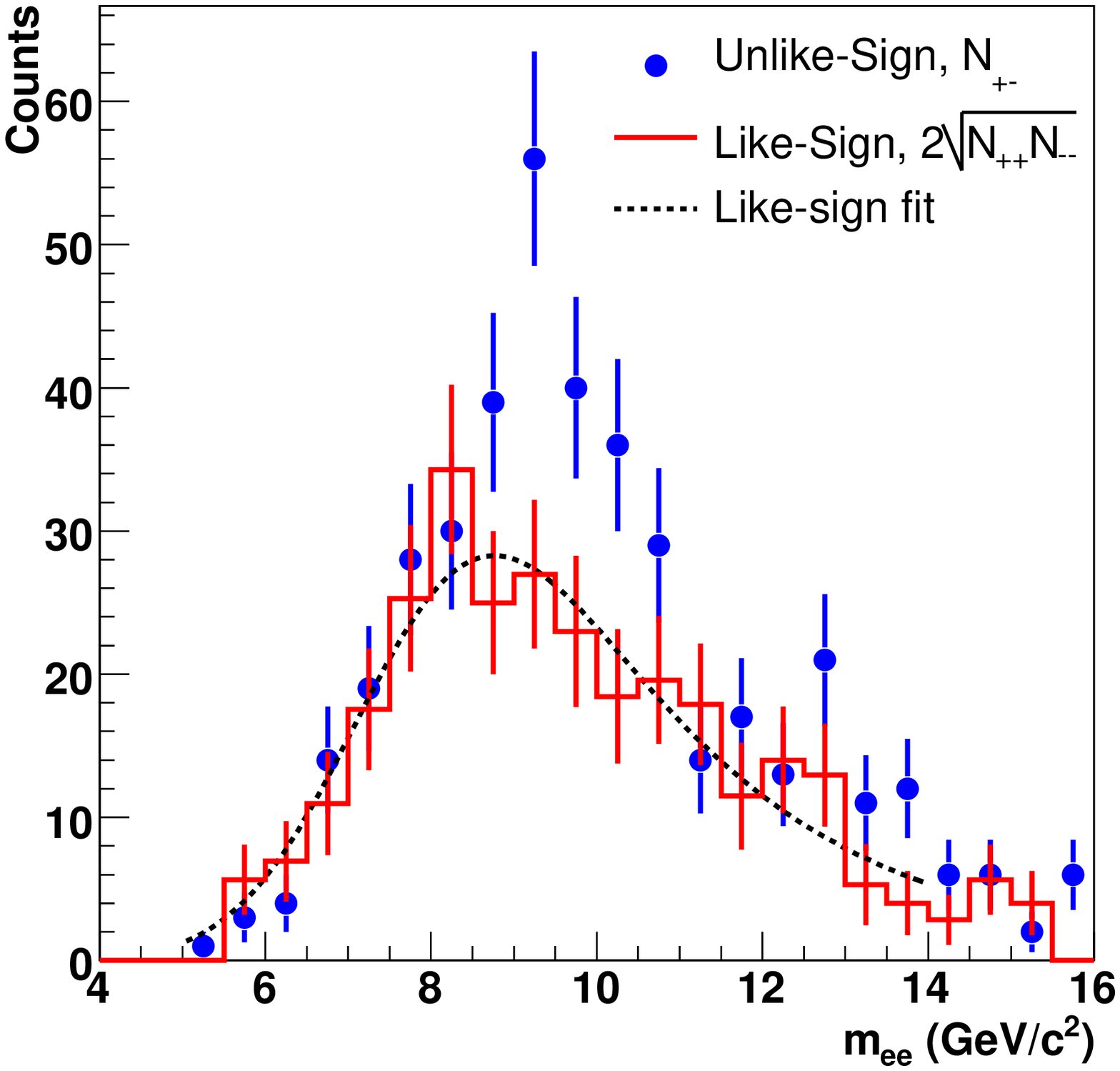}} 
   \resizebox{60mm}{!}{\includegraphics{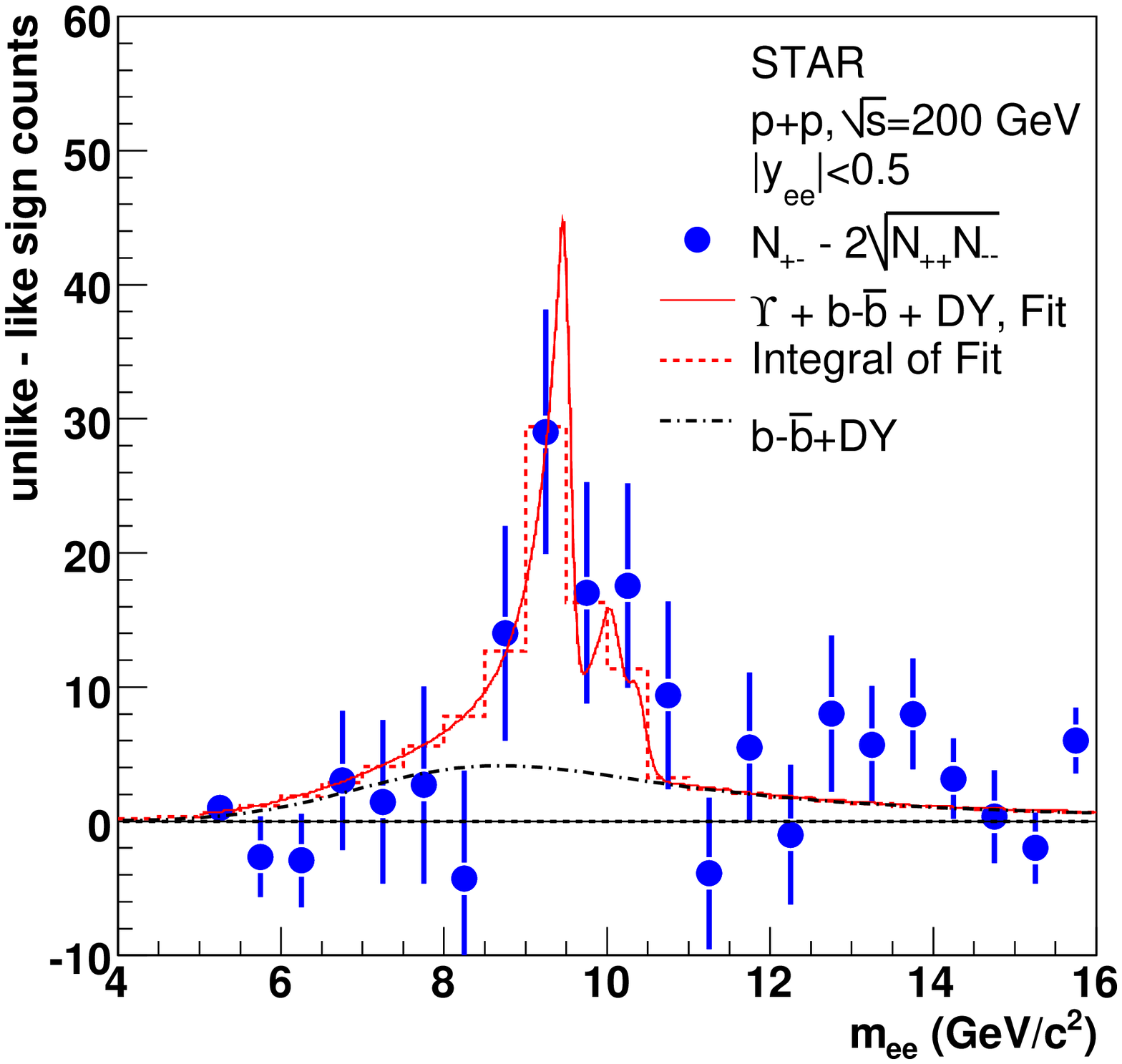}}
        \caption{
	(Left panel): Unlike-sign raw yield $N_{+-}$ in the region $|y_{ee}|\le$ 0.5 and like-sign combinatorial 
	background. 
	(Right panel): The $e^{+}e^{-}$ signal after subtracting the like-sign combinatorial background. 
	}
    \label{fig:invMass1}
   \end{center} 
\end{figure} 
\begin{figure}
\begin{center}
      \resizebox{65mm}{!}{\includegraphics{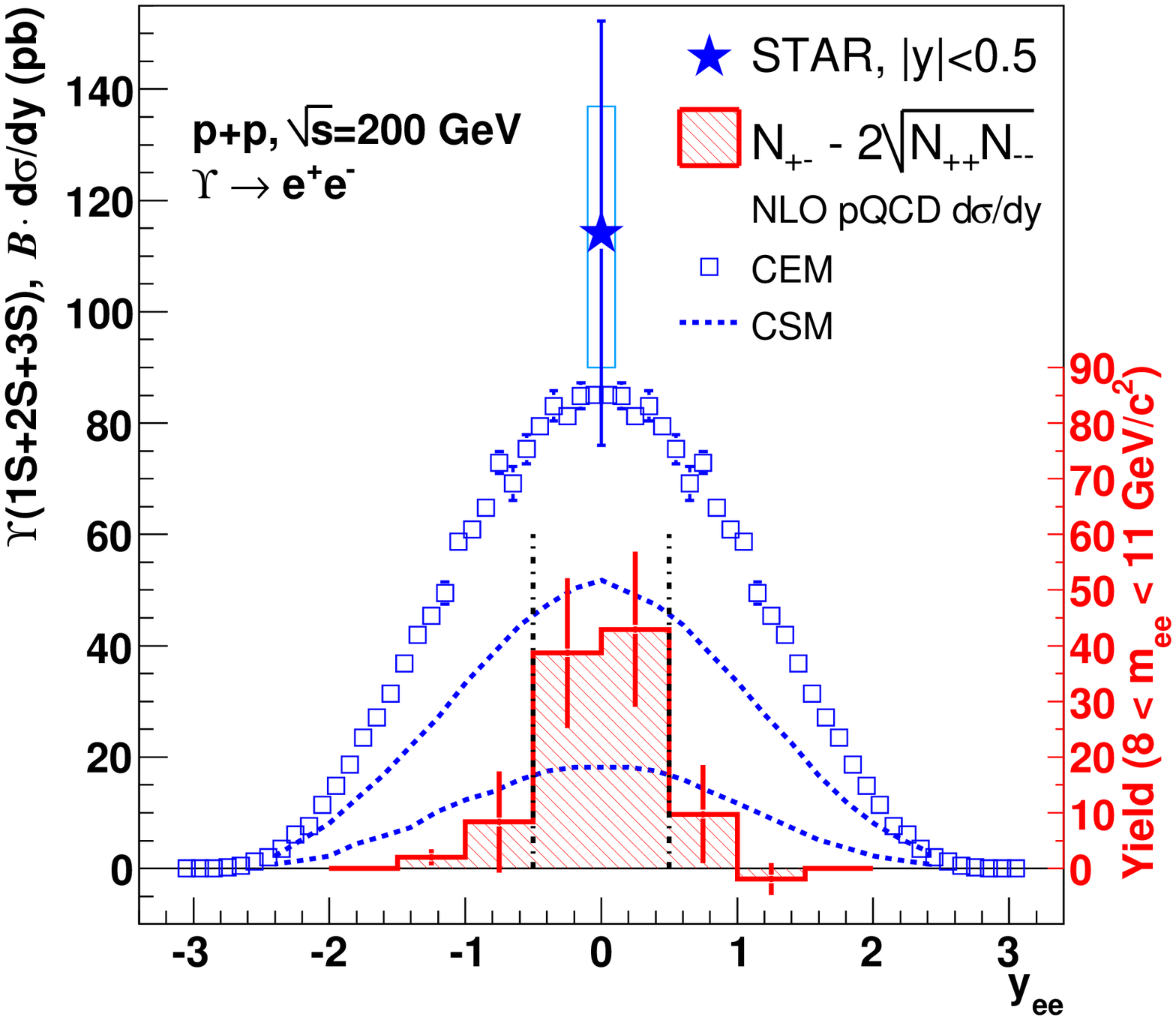}}
      \resizebox{59mm}{!}{\includegraphics{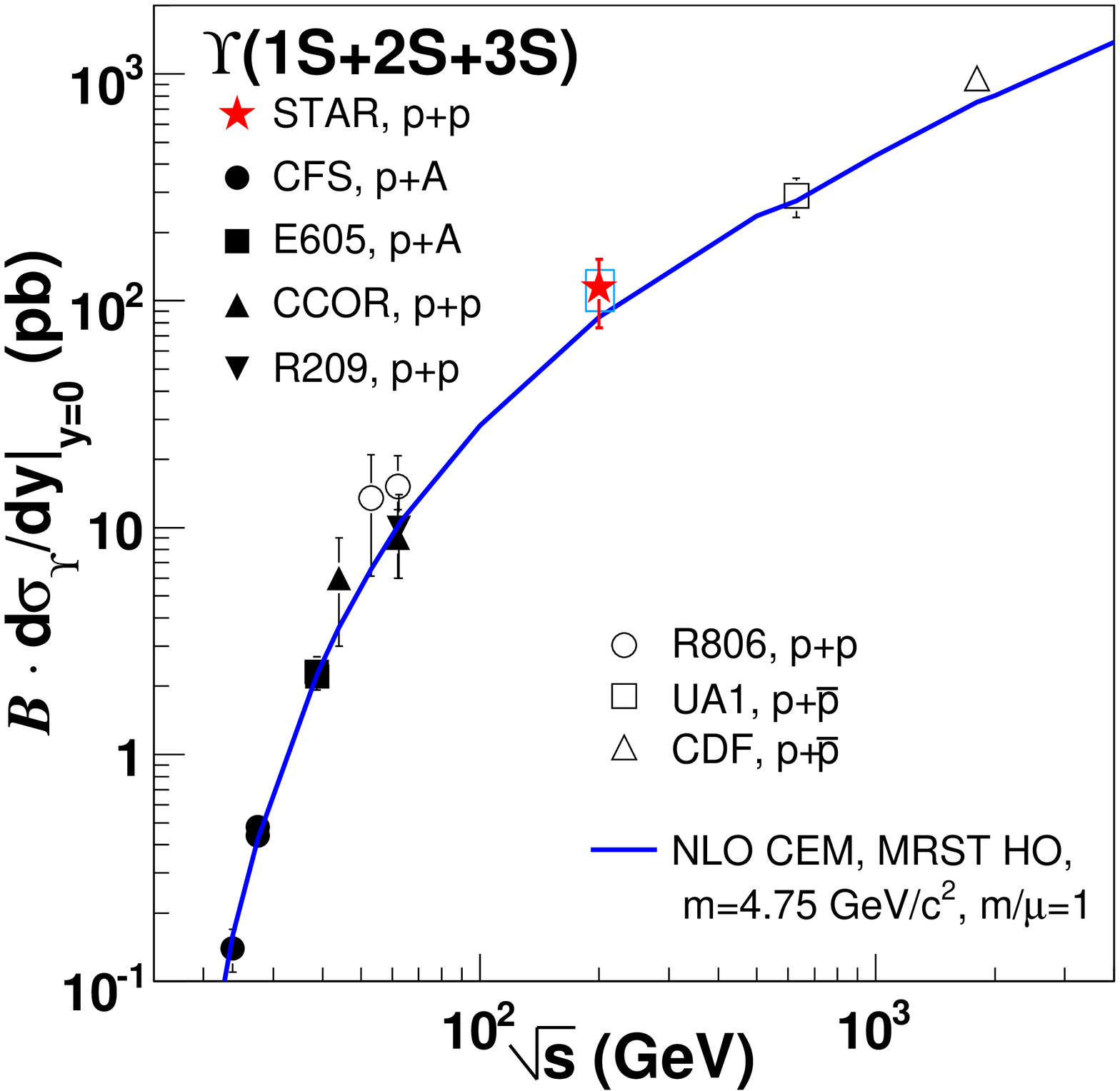}} 
        \caption{
	(Left panel): The STAR measurement of the midrapidity $\Upsilon(1S+2S+3S)$ cross section times 
	branching ratio into electrons.
	Error bars are statistical, the box shows the systematic uncertainty. The data is compared to theoretical predictions (see
	text).
	(Right panel): Evolution of the $\Upsilon(1S+2S+3S)$ cross section with $\sqrt{s}$ for the world data.
	}
    \label{fig:invMass2}
    \end{center}
\end{figure} 
\begin{figure}
\begin{center}
      \resizebox{65mm}{62mm}{\includegraphics{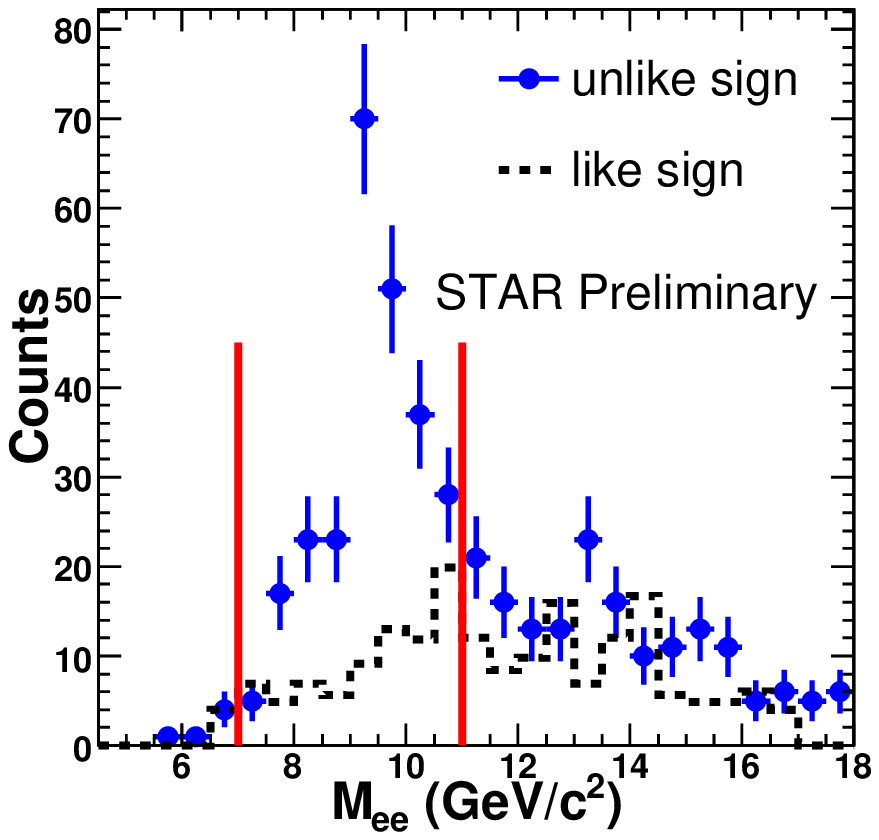}}
      \resizebox{65mm}{65mm}{\includegraphics{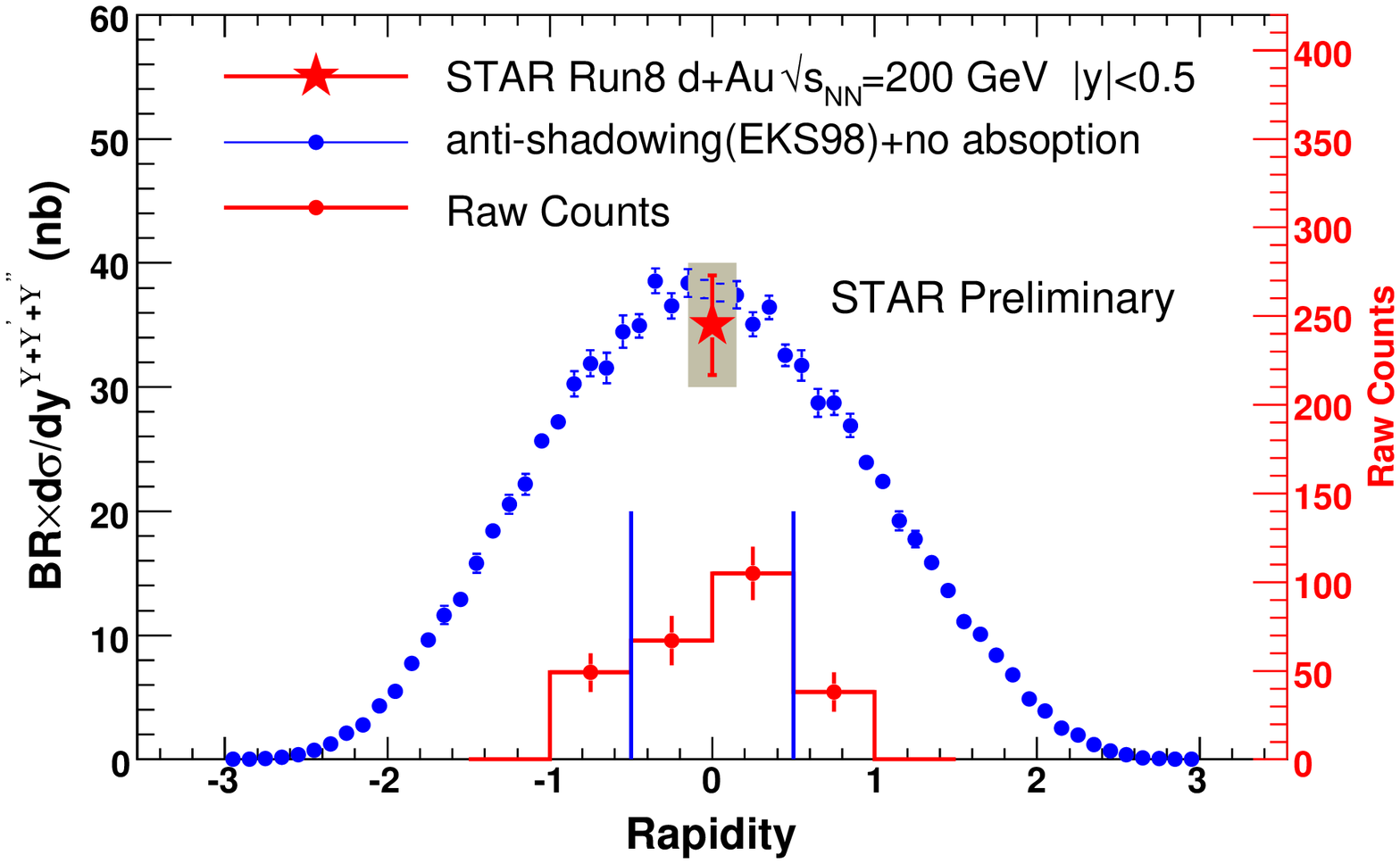}} 
        \caption{
	(Left panel): Unlike-sign raw yield $N_{+-}$ and background in d$+Au$ collisions. 
	(Right panel): The measured $\Upsilon(1S+2S+3S)$ cross section times 
	branching ratio into electrons in d$+Au$ collisions at midrapidity. The bar indicates the statistical error and the band
shows the systematic uncertainty. The cross section is compared to model prediction (see text). 
	}
    \label{fig:invMass3}
    \end{center}
\end{figure} 
\section{Results and Discussions}
The left panels of Fig.~\ref{fig:invMass1} and~\ref{fig:invMass3} show the unlike-sign and like-sign background 
invariant mass spectrum in $p+p$ collisions and in d$+Au$ collisions respectively. The total yield is found to be 
$N$ = 67 $\pm$ 22 (stat) in $p+p$ collisions and $N$ = 172 $\pm$ 20 (stat) in d$+Au$ collisions. 
The signal significance in d+Au ($\sim 8\sigma$) is greater than that in p+p ($\sim 3\sigma$) mainly due to the removal 
of the Silicon detectors in the former collisions which strongly reduced the photon conversions. 
Figure~\ref{fig:invMass1} (right panel) shows the data used to obtain the $\Upsilon$ and continuum yields. The data points are the
unlike-sign $e^{+}e^{-}$ signal after subtracting the like-sign combinatorial background. The fit includes the contributions from the
$\Upsilon(1S+2S+3S)$ states and the continuum contributions from Drell-Yan and $b\bar{b}$ shown as the solid-line function. 
The $\Upsilon(1S+2S+3S)$ cross section is found to be 
$\sum_{n=1}^3{\cal{B}}(nS)\times \sigma(nS)$ = 114 $\pm$ 38 (stat.) $^{+23}_{-24}$ (syst.) pb, and
the combined continuum contribution is estimated to be $(\sigma_{DY}+\sigma_{b\bar{b}})$ = 38 $\pm$ 24 pb in the integrated yield region.
Figure~\ref{fig:invMass2} (left panel) shows the acceptance in 
rapidity with the hashed histogram where the unlike-sign pairs are shown after the background 
subtraction.
The scale on the right
axis of the figure is used for the counts in the histogram. 
The measured cross section of the $\Upsilon(1S+2S+3S)$ is compared to the theoretical predictions 
of NLO CEM~\cite{CEM} and upper and lower bounds of NLO CSM~\cite{CSM}. 
The NLO CEM calculation is for $\Upsilon(1S)$ only and therefore the 
calculation is scaled with the branching ratios and the cross sections of the higher states in order to compare
with the measured cross section. The CSM calculation considers the direct $\Upsilon(1S)$ production only 
and therefore the calculation is divided by a factor of 0.42~\cite{CSM} to account for 
the feed-down from the P-states. 
While the NLO CEM is in agreement with the measured cross section, the upper bound of CSM 
underestimates the data by $\sim$ 2$\sigma$ effect.
Figure~\ref{fig:invMass2} (right panel) shows that the measured cross section of $\Upsilon(1S+2S+3S)$ 
is consistent with the world data trend
for similar measurements scanning a wide range of $\sqrt{s}$ 
in different collision systems. Figure~\ref{fig:invMass3} (right panel) shows the measured cross section 
of the $\Upsilon(1S+2S+3S)$ at midrapidity in d$+Au$ which is found to be 
$\sum_{n=1}^3{\cal{B}}(nS)\times \sigma(nS)$ =35 $\pm$ 4 (stat.) $\pm$ 5 (sys.) nb. 
The contribution from Drell-Yan and $b\overline{b}$ to the di-electron yield in the $\Upsilon$
mass region is estimated
to be $\sim$10$\%$ based on~\cite{CEM} and PYTHIA. The
detailed systematic uncertainty is under study. The measurement is consistent with NLO CEM calculation
that includes the
anti-shadowing effect,
and doesn't include absorption
effect. 
In order to study the cold nuclear matter effect, the nuclear modification factor, 
which is defined as the ratio of the cross section in d$+Au$ collisions to that in $p+p$ collisions scaled with the number of binary
collisions, is found to be $R_{dAu}$ = 0.98 $\pm$ 0.32 (stat.) $\pm$ 0.28 (sys.). 

\section{Conclusion}
The STAR experiment has measured the $\Upsilon(1S+2S+3S)\rightarrow e^{+}e^{-}$ cross section at mid-rapidity
in $p+p$ collisions and d$+Au$ collisions at $\sqrt{s}$ = 200 GeV. The measured value in $p+p$ is found to be consistent
with the world data trend as a function of $\sqrt{s}$. The NLO CEM calculations are in agreement with the measurements in
$p+p$ and d$+Au$ and the NLO CSM underestimated the measured cross section in $p+p$. The cold nuclear matter effect is not large,
however, more statistics 
is needed in order to quantify such effects. The presented measurements will be used as a baseline for
studying the hot nuclear matter effect in $Au+Au$ collisions. 
With increased luminosity, a separation between $\Upsilon$ states
will be possible at the STAR experiment, particularly after reducing the material budget 
by removing the inner tracker detectors.

\end{document}